\documentstyle[graphicx]{l-aa}
\begin{document}
\title{On Beaming due to Coherent inverse Compton Scattering}
\author{Thomas Kunzl\inst{1,3}, Harald Lesch\inst{1} and Axel
Jessner\inst{2}}
\institute{Institut f\"ur Astronomie und Astrophysik der Universit\"at
M\"unchen, Scheinerstr.1, D-81679 M\"unchen, Germany
\and
Max-Planck-Institut f\"ur Radioastronomie, Auf dem
H\"ugel 69, D-53121 Bonn, Germany
\and
Max-Planck-Institut f\"ur extraterrestrische Physik, Giessenbachstr.,
D-85740 Garching, Germany}

\offprints{Th. Kunzl,
          {\em tak@usm.uni-muenchen.de}}
\date{}
\maketitle

\begin{abstract}
We calculate the beaming of coherent emission by means of
geometrical optics. When spherical elementary waves are phase-coupled
in that way that constructive interference occurs along one axis, the
intensity is lower for off-axis viewing angles.
The spatial angle of constructive interference shrinks to a fraction of
 $4\pi$ when the length scale of the emission region is larger
 than the wavelength of the radiation.
 This leads to an  amplification of the received radiation in addition to the
relativistic beaming caused by the relativistic radial
outflow which is thought to  be important for the radio emission
of pulsars. The effect will be calculated  by numerical methods
and can be approximated by a simple analytic expression.
\end{abstract}

The main feature of a coherent process is that there are $N$ particles
radiating {\em in phase} (e.g. Melrose 1991 and references therein).
Therefore, the coherent
intensity of the radiation is calculated as $N^2$
times the single particle amplitude. In the case of $N$ incoherently
radiating particles the interference terms cancel and the total
intensity is given by $N$ times the square of the single particle amplitude.
Consequently the intensity of $N$ coherently radiating particles is $N$
times higher than in the incoherent case.

But, as will be shown, if the volume of coherently radiating particles
(hereafter coherence volume) is about the same size or larger than the
wavelength of the emitted radiation the phase coupling needed for the
increased intensities can only be maintained in one direction, i.e.
along one axis. Seen from off-axis angles there may even be
destructive interference leading to zero intensity. In this contribution we
calculate the angle at which the intensity drops to zero. Thus we get a
reasonable estimate for the beaming due to coherent radiation.

Assuming that the reason for coherence is a free electron maser (FEM)
process (i.e. inverse Compton scattering (ICS) of relativistic
electrons and nonlinear electrostatic plasma waves (Asseo et al. 1990;
Benford and Weatherall 1992; Asseo 1993)) we derive a formula for the
beaming angle depending only on the density, the Lorentz factor of the
electrons and the size of the coherence volume. 

Consider a volume of $N$ particles which are able to radiate in
phase. For simplicity we assume that the volume is a cube with one of
its axes in the direction where fully constructive interference
occurs. This seems reasonable since nonlinear electrostatic plasma waves
appear pancake-shaped due to Lorentz contraction (along the magnetic
field lines) and
the electrons are moving perpendicularly to the extension of these
plasma solitons.  

Therefore it seems natural to expect that the direction of
relative velocity between waves and particles describes the direction of
phase coupling (this shall be chosen to be the z- axis being parallel to
the vector (0,0,1)). The problem can be reduced to a two-dimensional one by
orientating the coordinate system in such a way that the observer is located
in the direction ($\sin\alpha, 0, \cos\alpha$) and  the y- axis can be disregarded
in the calculations.

If the coherence volume is $d^3$ (where $d$ is the length of a cube edge)
 we can choose the origin of the coordinate system to  be 
a corner such that in the cube all three coordinates are normalized
to the range of [0..$d$].

Throughout this paper we perform our calculations in the beam
frame, i.e. a Lorentz factor belongs to the nonlinear
wave structure approaching the electron beam (approximately equal to
the Lorentz factor of the beam seen by an observer).

The term {\em coherence volume} needs some clarification. It is the
volume occupied by the coherently radiating particles which is the
same as the volume of the  considered soliton. In other
words, we do not distinguish between the part of the soliton involved
in the reaction and the volume of beam electrons radiating coherently
except for  the Lorentz contraction since the volume is observed in the beam
frame. The soliton can be treated as the cause  of coherence whereas
the beam particles are coherently radiating when they interact with
the soliton. 

Next we calculate the phase shift of the wave from a radiating particle having the coordinates
$(x,z)$ relative to a similar particle at the origin (0,0) when radiating at 
an angle of $\alpha$. (The y-coordinate is disregarded due to the choice of the coordinates). The
phase difference can be expressed as
\begin{equation}
  \phi(x,z)=k\,(\delta(x,z)+z)
\end{equation}
where $k=2\pi/\lambda$ is the wave number of the emitted radiation,
$\delta(x,z)$ denotes the difference in pathlengths for radiation from the two particles
(where the usual approximations are adopted for calculations of interference
at optical grids). The second term inside the brackets, $z$,
describes the {\em intrinsic} phase difference which is necessary for
constructive interference in z-direction. Obviously a particle which
is right ahead of another in the distance of $z$ must have a phase
delay of $kz$ to interfere constructively.

\begin{figure}[htbp]
  \centerline{\includegraphics*[width=\linewidth]{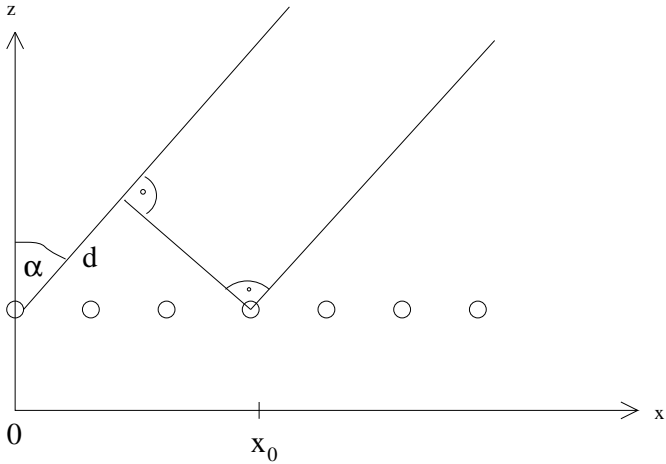}}
  \caption{Schematic picture showing how the path difference for two
  points is calculated having the same z- coordinate.}
\end{figure}
\begin{figure}[htbp]
  \centerline{\includegraphics*[width=\linewidth]{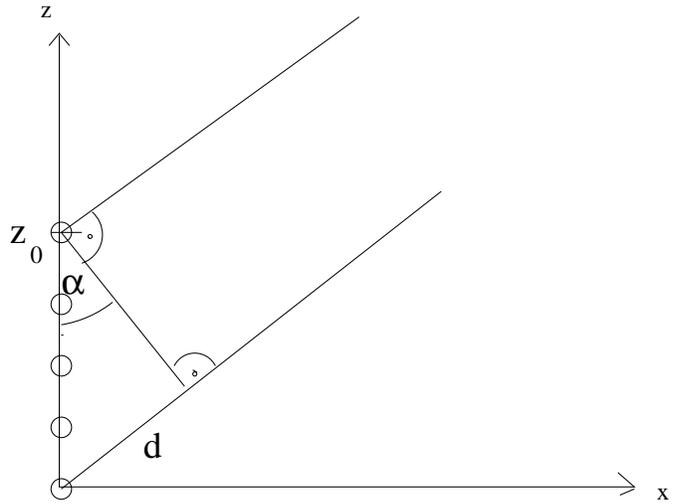}}
  \caption{The same as Fig. 1, but for two particles having the same x-
  coordinate.}
\end{figure}
$\delta_0$ is now calculated for two special cases:

1) z=0: It can be seen from Fig. 1 that $\delta_0=x\sin\alpha
        \approx x\alpha$ for small values of $\alpha$.

2) x=0: In this case we find $\delta_0=-z\cos\alpha$ (cf. Fig. 2)
   and thus $\phi(0,z)=z\,(1-\cos\alpha)\approx z\alpha^2/2$ for small
   $\alpha$

We add these two phase differences to get the general
result. This renders:
\begin{equation}
  \phi(x,z)=\frac{2\pi}{\lambda}\left(x\alpha+\frac{1}{2}\,z\alpha^2\right)
\end{equation}

For calculating the total intensity relative to the maximum value we
integrate the amplitudes over one phase and over  the coherence
volume and  take the square of the integral.
In principle one has to sum up the complex amplitudes of waves from all the particles
but replacing the sum by an integral does not affect the results
heavily when $N\gg 1$. Using the integral has the advantage that  particles need not to be arranged in a
grid (as the electron temperature might be quite high) and the result will become more representative
of the physics involved.

It is reasonable to assume that coherence is not provided
instaneously at its maximum possible number $N$, but that the coherence
volume will be zero initially and then expand in some way.
Since we are mainly interested in the {\em maximum} possible effect, we do
not assume a certain time dependence but just set a maximum size of
the coherence cell. The coherence is caused by a plasma process,
therefore the most natural assumption is to set the size to $d=c/\nu_{\rm
pe}$. Here $\nu_{\rm pe}=\sqrt{(ne^2)/(m_{\rm e}\varepsilon_0\gamma)}/(2\pi)$
is the (frame invariant) plasma
frequency (cf. Weatherall and Benford, 1991).

By re-scaling the integration interval to [0..1]
(that means transforming the integration variables to dimensionless
lengths in units of the coherence cell extension)
we finally obtain for the relative phase averaged wave amplitudes  in the direction of
$\alpha$
\begin{equation}
\frac{A}{A_0}=\int\limits_0^1\,\int\limits_0^1\,\cos
\left[\frac{2\pi c}{\lambda\nu_{\rm pe}}\left(x\alpha+\frac{1}{2}z\alpha^2
\right)\right]\,{\rm d}z\,{\rm d}x\,
\end{equation}
where $A$ is the amplitude of the wave and $A_0=A(\alpha=0)$ the corresponding amplitude 
of waves emitted parallel to the particle propagation.

The double integral simplifies to
\begin{equation}
  \frac{A}{A_0}=\frac{2}{M^2\alpha^3}\left(\sin a\sin b-\cos a\cos b
  +\cos a+\cos b-1\right)
\end{equation}
where $M=2\pi\,c/(\lambda\nu_{\rm pe})$, $a=M\alpha,\;b=M\alpha^2/2$.
The first  zero of the above equation yields the angle of
the minimum and we find the solution $a=\pi-b$ which leads to a simple
quadratic equation for $\alpha$. The final
result then reads
\begin{equation}
  \alpha_{\rm min}=\sqrt{1+\frac{2\pi}{M}}-1\approx
  \frac{\lambda\nu_{\rm pe}}{2c}\,.
\end{equation}
The last approximation is valid as long as $M\gg1$ which is always 
the case as we consider only plasma processes for radio
emission, and thus  can write $c/(\lambda\nu_{\rm pe})=\nu_{\rm
em}/\nu_{\rm pe}=\gamma^\varsigma$ where $\varsigma$ is an emission
model dependent parameter which is at least 1.

This result can be compared to a more exact geometric model that we
obtain by numerical integration over the phases of waves  emitted from a soliton.

The soliton shape is chosen since a coherent plasma process requires
efficient bunching of the radiating particles. Physically speaking this is the same as a
high density modulation or a strong longitudinal plasma wave, which is well described
by an almost non-dispersive longitudinal plasma soliton
with a certain extension perpendicular to the magnetic field lines.

We assume a cylindrically symmetric shape like
${\rm cosh}\left(\frac{\rho}{R}\right)^{-2}$ (see Fig. 3)
where $\rho$ denotes the distance from the center of
the soliton in units of its characteristic length 
(called "extension" of the soliton in the simplified discussion
before) and $R$ as the characteristic length scale equal to
$c/\nu_{\rm pe}$ or equal to $d$ used in the simple approximation
above (e.g. Weatherall 1997).

\begin{figure}[htbp]
  \centerline{\includegraphics*[width=\linewidth]{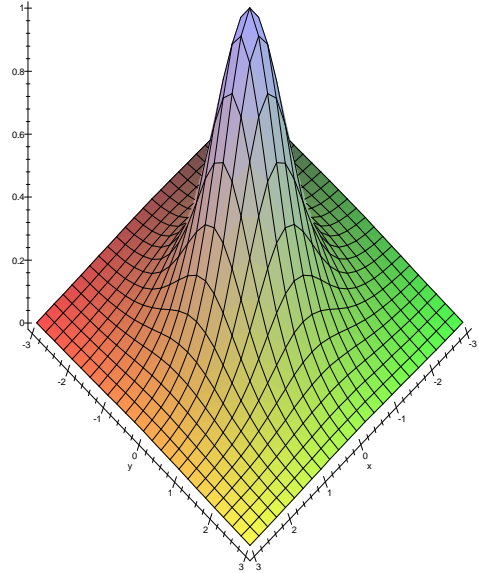}}
  \caption{Typical shape of a soliton. The maximum amplitude of the
  soliton is normalized to 1,  
  and points in radial direction (z). The plane spanned by the line of sight
  and the direction of the outflowing particles as assumed to be the
  y=0 plane}. X- and y- axis are scaled in units of the characteristic
  length (see text).
\end{figure}

 The solitons developing from electrostatic plasma
waves are negligibly thin in the direction of the field lines but have a large  perpendicular extension.
Let the observer be placed at a large distance $s$ and at an angle  $\alpha$ with respect to the
normal direction.
 Again we
assume that the positive x- axis is the projection of the observer's
line of sight. Then, $s$ being the distance and $\phi$ the meridional
angle in the soliton plane we find the expression 
\begin{equation}
  r^2=s^2-2s\rho\,\sin\phi\,\cos\alpha+\rho^2.
\end{equation}
A spherical wave emitted from a point source can be written as
\begin{equation}
  \Psi(r)=\frac{{\rm e}^{-{\rm i}kr}}{r}
\end{equation}
Thus the superposition of all elementary waves from a soliton reads
\begin{equation}
  A_{\rm sol}(s,\theta)=A_0\int\limits_0^\infty \int\limits_0^{2\pi}
  \frac{{\rm e}^{-{\rm i}kr(s,\rho,\alpha,\phi)}}{r(s,\rho,\alpha,\phi)}
  \rho{\rm cosh}\left(\frac{\rho}{R}\right)^{-2}\,
  {\rm d}\phi\,{\rm d}\rho.
\end{equation}

\begin{figure}[htbp]
   \centerline{\includegraphics*[width=0.8\linewidth]{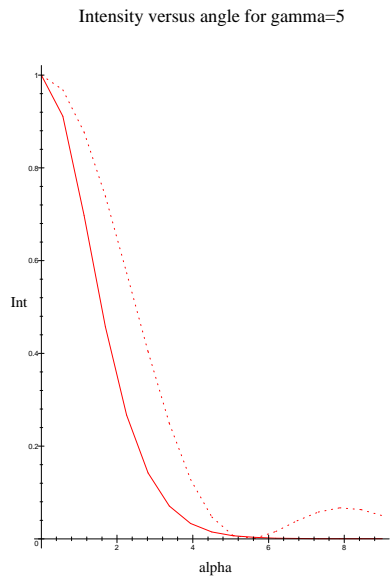}}
   \caption{Beaming pattern for the coherent emission. The exact result
   is shown by the fully drawn line whereas the dotted line shows the
   approximation. The angle between the line of sight and the direction
   of movement is given in degrees and the intensity is normalized to 1
   for $\alpha=0$}
   \label{Vergleich}
\end{figure}

$k$ is the wave number with the same definition as above.
This equation can be treated numerically. We find that the result
is not altered significantly but only slightly modified by the
different shape of the coherence cell. As can be seen in Fig.
(\ref{Vergleich}) the beaming is even stronger than found in the
simple approximation.
\footnote{ No great discrepancy arises even though we
treated the "coherence cell" as a cubic box  in the first case  and in the numerical
calculations we assumed the soliton to be pancake-shaped. 
But if the phase-coupled emitter extends over many wavelegths in all directions,
 $\lambda/d$ will dominate the interference effects and  the symmetry of the
emitter will be of secondary importance. }

As mentioned above we assume a plasma process to be the origin of
coherent emission. As an illustrative example we take the free
electron maser mechanism which can ({\em cum grano salis})
be treated as inverse Compton scattering of plasma solitons and
relativistic electrons. Then the incoming electron sees a plasma wave
with a wavelength that is equal to that measured in the soliton frame
(which is almost the same as the rest frame for very low-relativistic
solitons) divided by $\gamma$ due to relativistic space
contraction. 
Seen in the beam frame the electrostatic wave is 
simply reflected with the same frequency. In the rest frame the
reflected wave has a frequency that is higher by another factor of $\gamma$
and is given by $\gamma^2\omega_{\rm pe}$ as can  be verified by
a Lorentz transformation from the beam frame to the observer's frame.
  A wave of this frequency cannot
propagate as a plasma wave any more and will be decoupled from the
plasma  propagating as a free electromagnetic wave. Because
 we performed our calculations in the beam frame,  only one
$\gamma$ is required in the formula for the emitted frequency; that means
$\varsigma=1$ in the formula given above.
%

Thus we finally find the expression
\begin{equation}
  \label{beaming}
  \alpha_{\rm min}=\frac{1}{2\gamma}
\end{equation}
From that equation we obtain that the beam becomes sharper
when $\gamma$
is larger. Because we require the strongest coherence at {\em
low} frequencies (and therefore naturally low Lorentz factors) it is
natural to assume that we have non-stationary processes which start
incoherently (at high Lorentz factors) and become more and more
coherent when $\gamma$ decreases. Equation (\ref{beaming}) should
only be valid for the lowest Lorentz factors when coherence is fully
developed.

The angle of the minimum now provides an estimate for the spatial
beaming angle due to coherence. This is given by
\begin{equation}
  \Delta\Omega=\pi\alpha_{\rm min}^2
\end{equation}
which is true for small angles. This corresponds to a {\em relative
beaming factor} of
\begin{equation}
  \eta_1=\frac{4\pi}{\Delta\Omega}=16\gamma^2
\end{equation}
in the beam frame.

For a fixed observer in addition to coherent beaming there is, of
course, beaming by a factor
of $4\gamma^2/\pi$ (Rybicki and Lightman 1979)
due to the relativistic motion of the solitons themselves. Thus an
observer sees a relative beaming factor that reads
\begin{equation}
  \eta=\frac{4\pi}{\Delta\Omega}\gamma^2=
  \frac{64 \gamma^4}{\pi}
\end{equation}

It has been shown above that in coherent processes one can find a
natural beaming effect that is caused by the non-isotropic emission
pattern of a phase-coupled grid of point sources due to interference.
This leads to higher fluxes of coherent radiation in special
directions which can be important for estimations of brightness
temperatures or observed radiation "power" (that means the power that
would have to be emitted if the radiation was isotropic in the
observers frame, see Manchester and Taylor 1977). The described effect
might be a contributional cause of the very intense fluxes of radio
photons in pulsars. Some detailled discussion of the flux problem will
be subject of a later paper.

\end{document}